\documentclass[sigconf]{acmart}
\usepackage{CJKutf8}
\usepackage{amsmath,latexsym,amsfonts,amsthm,cleveref,bm,bbm}
\usepackage{stmaryrd}
\usepackage{float}
\usepackage{balance}

\AtBeginDocument{%
  \providecommand\BibTeX{{%
    \normalfont B\kern-0.5em{\scshape i\kern-0.25em b}\kern-0.8em\TeX}}}

\copyrightyear{2024}
\acmYear{2024}
\setcopyright{acmlicensed}\acmConference[KDD '24]{Proceedings of the 30th ACM SIGKDD Conference on Knowledge Discovery and Data Mining}{August 25--29, 2024}{Barcelona, Spain}
\acmBooktitle{Proceedings of the 30th ACM SIGKDD Conference on Knowledge Discovery and Data Mining (KDD '24), August 25--29, 2024, Barcelona, Spain}
\acmDOI{10.1145/3637528.3671559}
\acmISBN{979-8-4007-0490-1/24/08}

\begin{document}

\title{Deep Bag-of-Words Model: An Efficient and Interpretable Relevance Architecture for Chinese E-Commerce}

\author{Zhe Lin}
\email{linzhe.lin@taobao.com}
\affiliation{%
  \institution{Alibaba Group}
  \city{HangZhou}
  \country{China}
}

\author{Jiwei Tan}
\authornote{Jiwei Tan is the corresponding author.}
\email{jiwei.tjw@taobao.com}
\affiliation{%
  \institution{Alibaba Group}
  \city{HangZhou}
  \country{China}
}
\author{Dan Ou}
\email{oudan.od@taobao.com}
\affiliation{%
  \institution{Alibaba Group}
  \city{HangZhou}
  \country{China}
}

\author{Xi Chen}
\email{gongda.cx@taobao.com}
\affiliation{%
  \institution{Alibaba Group}
  \city{HangZhou}
  \country{China}
}

\author{Shaowei Yao}
\email{yaoshaowei@taobao.com}
\affiliation{%
  \institution{Alibaba Group}
  \city{HangZhou}
  \country{China}
}

\author{Bo Zheng}
\email{bozheng@alibaba-inc.com}
\affiliation{%
  \institution{Alibaba Group}
  \city{HangZhou}
  \country{China}
}

\renewcommand{\shortauthors}{Zhe Lin, et al.}

\begin{abstract}
Text relevance or text matching of query and product is an essential technique for the e-commerce search system to ensure that the displayed products can match the intent of the query. Many studies focus on improving the performance of the relevance model in search system. Recently, pre-trained language models like BERT have achieved promising performance on the text relevance task. 
While these models perform well on the offline test dataset, there are still obstacles to deploy the pre-trained language model to the online system as their high latency. 
The two-tower model is extensively employed in industrial scenarios, owing to its ability to harmonize performance with computational efficiency. Regrettably, such models present an opaque ``black box'' nature, which prevents developers from making special optimizations. In this paper, we raise \textbf{deep} \textbf{B}ag-\textbf{o}f-\textbf{W}ords (\textbf{DeepBoW}) model, an efficient and interpretable relevance architecture for Chinese e-commerce. Our approach proposes to encode the query and the product into the sparse BoW representation, which is a set of word-weight pairs. The weight means the important or the relevant score between the corresponding word and the raw text. The relevance score is measured by the accumulation of the matched word between the sparse BoW representation of the query and the product. 
Compared to popular dense distributed representation that usually suffers from the drawback of black-box, the most advantage of the proposed representation model is highly explainable and interventionable, which is a superior advantage to the deployment and operation of online search engines.
Moreover, the online efficiency of the proposed model is even better than the most efficient inner product form of dense representation. 
The proposed model is experimented on three different datasets for learning the sparse BoW representations, including the human-annotation set, the search-log set and the click-through set. Then the models are evaluated by experienced human annotators. Both the auto metrics and the online evaluations show our DeepBoW model achieves competitive performance while the online inference is much more efficient than the other models. Our DeepBoW model has already deployed to the biggest Chinese e-commerce search engine Taobao and served the entire search traffic for over 6 months.
\end{abstract}





\maketitle

\section{Introduction}

The popularization of mobile internet has significantly elevated the prominence of online commerce in daily life. Hundreds of millions of customers purchase products they want on large e-commerce portals, such as Taobao and Amazon. The search engine emerges as the essential technology in assisting users to discover products that are in accord with their preferences. Different from general search engines like Google, commercial e-commerce search engines are usually designed to improve the user's engagement and conversion, possibly at the cost of relevance in some cases \cite{10.1145/3336191.3371780}. The exhibition of products in search results that are inconsistent with the query intent has the potential to diminish the customer experience and hamper customers’ long-term trust and engagement. Consequently, measuring relevance between the text of search query and products to filter the irrelevant products is an indispensable process in the e-commerce search engine.

Text relevance has been a long-standing research topic due to its importance in information retrieval and the search engine. Researchers and engineers have been dedicated to the pursuit of an efficient and robust model to accurately measure the text relevance between the query and the product in the e-commerce scenario. Conventional methodologies have traditionally harnessed attributes such as the word matching ratio, Term Frequency-Inverse Document Frequency (TF-IDF, notably BM25), or cosine similarity to serve as the relevance score, often yielding strong baseline performances. Nevertheless, these word-matching approaches may cause inaccuracies due to inconsistent linguistic expressions of identical meanings such as synonyms. This issue is particularly pronounced in the e-commerce scenario since the queries are usually described by users in daily language while the products are described by sellers in professional language. Thus severe vocabulary gap may exist between queries and products \cite{10.1145/3289600.3291039}.


With the development of deep learning technology, neural models have shown their advantage in addressing the semantic matching problem for the text relevance task \citep{NIPS2014_b9d487a3,10.5555/3060832.3061030,10.5555/3016100.3016292}.
Recently, pre-trained language models like BERT \cite{devlin-etal-2019-bert} achieve excellent results in various NLP tasks including text matching. 
Unfortunately, typical paradigm of the BERT-based relevance model is the interaction-based structure, which needs to encode the query-document pair in real time to measure their relevance. This makes it difficult to be deployed to  online systems with large traffic due to the high computation and latency.  
Consequently, it is usually impractical to deploy the pre-trained model directly to search systems. 
To address this problem, the representation-based model, also known as the two-tower model, is proposed and mostly applied to industrial search systems. It usually pre-computes the embeddings of query/document respectively, and measures the relevance online from the embeddings. SiameseBERT \cite{reimers-2019-sentence-bert} leverages BERT as the encoder and calculates the cosine similarity between the dense embeddings of query and document as the relevance score. Some  studies like ReprBERT \cite{10.1145/3534678.3539090} explore the more complex MLP classifier to compute the relevance score between two dense embeddings, which can achieve improved performance.

However, the representation-based model with dense embeddings still faces two major issues. First, the dense embedding may lose the detailed semantic information of the text, especially for low-frequency words like product models, entity names, or even brand identifiers. These words are essential in the e-commerce relevance task. Second, the dense embedding presents an opaque ``black box'' nature, which prevents developers from comprehensively understanding the model's methodology for calculating relevance scores. Developers often find it difficult to analyze the reasons for bad cases in the online system and implement targeted optimizations. 
In contrast, traditional word-matching algorithms like BM25 continue to be favored in numerous industrial applications\cite{thakur2021beir} due to their high efficiency and robust interpretability. Such word-based algorithm can capture the match of words that are low-frequency but essential for the text-relevance task. Unfortunately, these methods are not without their constraints. They fall short in recognizing different linguistic expressions that convey identical meanings, such as synonyms, thereby limiting their effectiveness in semantic matching tasks.

Is it possible to combine the advantages of both deep semantic models and word matching methods? In this paper, we propose  \textbf{Deep} \textbf{B}ag-\textbf{o}f-\textbf{W}ords (\textbf{DeepBoW}), which can leverage the pre-trained language model with large language corpora to improve semantic modeling while preserving the computational efficiency and interpretability of the word-matching method. We realize this by designing to learn sparse bag-of-words representation through deep neural networks. Our model generates the query/product 
high-dimensional representation (called the BoW representation) instead of the low-dimensional distributed representation (i.e. embedding). The dimensional size is the same as the size of the vocabulary. Each position in this high-dimensional representation corresponds to a word in the vocabulary, and with a value represents the weight of this word in the BoW representation, just like the BoW vector of TF-IDF.
The proposed DeepBoW model encounters two predominant challenges. Firstly, due to the opaque nature of neural network models, often colloquially referred to as "black boxes", it is challenging to correlate positions within high-dimensional representations to specific words in the vocabulary. Secondly, the vocabulary size is usually much larger than the dimension of dense embedding. Expanding the dimensional size to the vocabulary size may explode the computation and storage resources.
For the first challenge, we elaborately design the architecture of the model and loss function to align the position of the high-dimensional representation and the corresponding word in the vocabulary. For the second challenge, we add a sparse constraint in the loss function to reduce valid positions in the high-dimensional representation since the query/product should not include all words of vocabulary. Finally, we sample the high-dimensional representation to a small set of non-zero word-weight pairs, which is named as sparse BoW representation. 
In addition, although queries and product descriptions in e-commerce primarily consist of keywords, there still exists some semantic dependency on the word combination, which unigrams may not capture adequately. For example, brand names with multi-words may be incorrectly matched if they are separated. Consequently, we propose to model n-gram in our DeepBoW model, meanwhile introducing an n-gram hashing vocabulary strategy to avoid the explosion of vocabulary size. Finally, with the sparse BoW representation, the relevance score is measured in a most easy way as the weight accumulation of the matching words in the query/product's sparse BoW representations, which makes it highly efficient and interpretable.

The proposed DeepBoW model is evaluated on the three industrial datasets. The results show that our DeepBoW model achieves more than 2.1\% AUC improvement compared to the state-of-the-art two-tower relevance model on Chinese e-commerce. The sparse BoW representation generated by our DeepBoW model has positive interpretability and supports easy problem investigation and intervention for online systems. The time complexity of the online relevance score computing program can be optimized to $\mathcal{O}(N)$ by leveraging the two-pointer algorithm, which is faster than previous state-of-the-art relevance model.
The contribution of this paper is summarized as follows:

\begin{itemize}

    \item We reveal that the semantics of query and product in the e-commerce scenario can be represented by the bag-of-words vectors with importance weight. We show the BoW representation can be more suitable for the industrial e-commerce search system as it's more interpretable and flexible than the dense embedding from the BERT-based model. 

    \item  We introduce an innovative architecture designed to encode query/product into two distinct sparse Bag-of-Words (BoW) representations. We elucidate a methodology by which relevance scoring, based on these sparse BoW representations, can reduce the latency of the online relevance model while preserving competitive performance.

    \item Our proposed DeepBoW model is evaluated both on offline human-annotation datasets in Chinese and online A/B testing, achieving strong performance and efficiency. The model has already been deployed on the largest Chinese e-commerce platform Taobao, and has been serving the entire search traffic for over six months.
    
\end{itemize}

The rest of this paper is organized as follows. In Section \ref{related_work} we introduce related work. The proposed method is detailed in Section \ref{sec:method}, and experimental results are presented in Section \ref{sec:experiments}. Finally, in Section 5 we conclude this work and discuss the future work.

\section{Related Work}
\label{related_work}

\subsection{Text Matching}

The text matching task takes textual sequences as input and predicts a numerical value or a category indicating their relationship. Text matching is a long-stand problem and a hot research topic as it's important in information retrieval and search system. The e-commerce relevance learning can be regarded as a text-matching task. Early work mostly performs keyword-based matching that relies on manually defined features, such as TF-IDF similarity and BM25 \cite{robertson-1994-okapi}. These methods cannot effectively utilize raw text features and usually fail to evaluate semantic relevance.

Recently with the development of deep learning, neural-based text-matching models have been employed to solve the semantic matching task and have achieved promising performance. The architecture of the neural-based text-matching model can be roughly divided into the interaction-based model and the representation-based (two-tower) model. The interaction-based model \cite{NIPS2014_b9d487a3,10.5555/3016100.3016292,10.5555/3016100.3016292,10.5555/3060832.3061030,parikh-etal-2016-decomposable} usually puts all candidate text together as the input. The model can employ the full textual feature to calculate the matching feature as the low layer, and then aggregate the partial evidence of relevance to make the final decision. So interaction-based model can leverage sophisticated techniques in the aggregation procedure and achieve better performance. 
More recent studies are built upon the pre-trained language model like BERT \cite{devlin-etal-2019-bert}. With an extremely large corpus for pre-training, these methods can achieve new state-of-the-art performance on various benchmarks. The architecture of these models is the pre-trained bidirectional Transformer \cite{NIPS2017_transformer}, which can also be regarded as an interaction-based model. The typical paradigm of the BERT-based relevance model is to feed text pair into BERT and then build a non-linear classifier upon BERT’s [CLS] output token to predict the relevance score \cite{nogueira2019multistage,nogueira2020passage}. 

Although having excellent performance in the text-matching task, interaction-based models are still hard to be deployed to practical online service as they are mostly time-consuming, and the features of queries and documents cannot be pre-computed offline. Two-tower models are widely used in many online search systems. The two-tower model consists of two identical neural networks, each taking one of the two inputs. DSSM \citep{10.1145/2505515.2505665,10.1145/2661829.2661935} is a two-tower model that employs two separate deep full-connected networks to encode the candidate texts. Meanwhile, more sophisticated architectures can be adopted to enhance the ability of learning semantic representations. 
LSTM-DSSM \cite{palangi2015semantic} and LSTM-RNN \cite{10.1109/TASLP.2016.2520371} use RNNs to explicitly model word dependencies in the sentences. Typically dot-product, cosine, or parameterized non-linear layers are used to measure the similarity between representations of all candidate texts. Since individually encoding both the queries and documents, the embeddings of them can be pre-computed offline. Therefore, representation-based methods are online efficient and are widely used in industrial search engines. However, the encoding procedure of two inputs is independent with each other, making the final classifier hard to predict their relationship.

\subsection{Search Relevance Matching}

Relevance in search engine is a special sub-task of the text-matching which computes the relevance score between the query and the product (as the document). Different from the typical text-matching task which all input texts are semantically similar and homogeneous (i.e. having comparable lengths), the length of query may be much shorter than the length of document. Query only needs to match the partial semantics in the document. 

A large number of models are proposed for conducting matching in search. Neural Tensor Network (NTN) \cite{10.5555/2999611.2999715} is originally proposed to explicitly model multiple interactions of relational data. It achieves powerful representation ability that can represent multiple similarity functions, including cosine similarity, dot product, and bilinear product, etc. 
\citet{qiao2019understanding} apply the BERT model to ad-hoc retrieval and passage retrieval. \citet{reimers-gurevych-2019-sentence} propose Sentence-BERT for reducing the computational overhead for text matching. 
\citet{Bai2020SparTermLT} conduct a pilot study to map the frequency-based and BoW representation of a document to a
sparse term importance distribution for text retrieval.

E-commerce search is a special scenario of the Web search system. Both tasks model the semantic matching between query and candidate and require high efficiency and low latency in the online search system. Differently, in Web search the query and document are usually very different in length, making most methods not feasible for the e-commerce relevance task. 
Currently, there is not a commonly-used public benchmark for the Chinese e-commerce relevance task, so previous works usually evaluate their models on the online service and the real-world dataset constructed from the online platforms. 
\citet{10.1145/2983323.2983769} introduce a typical framework for e-commerce relevance learning. A Siamese network is adopted to learn pair-wise relevance of two products to a query. They investigate training the model with user clicks and batch negatives, followed by finetuning with human supervision to calibrate the score by pair-wise learning. 
\citet{10.1145/3289600.3291039} propose a co-training framework to address the data sparseness problem by investigating the instinctive connection between query rewriting and semantic matching. 
\citet{10.1145/3442381.3450129} propose to learn a two-tower relevance model from click-through data in e-commerce by designing a point-wise loss function.
\citet{Zhang2019ImprovingSM} also find the weakness of training with click signals, and address this problem by proposing a multi-task learning framework of query intent classification and semantic textual similarity to improve semantic matching efficiency. \citet{10.1145/3292500.3330759} introduce a 3-part hinge loss to differentiate multiple types of training data. They classified training instances into three categories: random negative examples, impressed but not purchased examples, and purchased items. Recently \citet{10.1145/3534678.3539090} propose ReprBERT, which has the advantages of both excellent performance and low latency, by distilling the interaction-based BERT model to a representation-based architecture. This framework is taken as the baseline of our model.

\begin{figure*}[htb]
\centering
\includegraphics[scale=0.38]{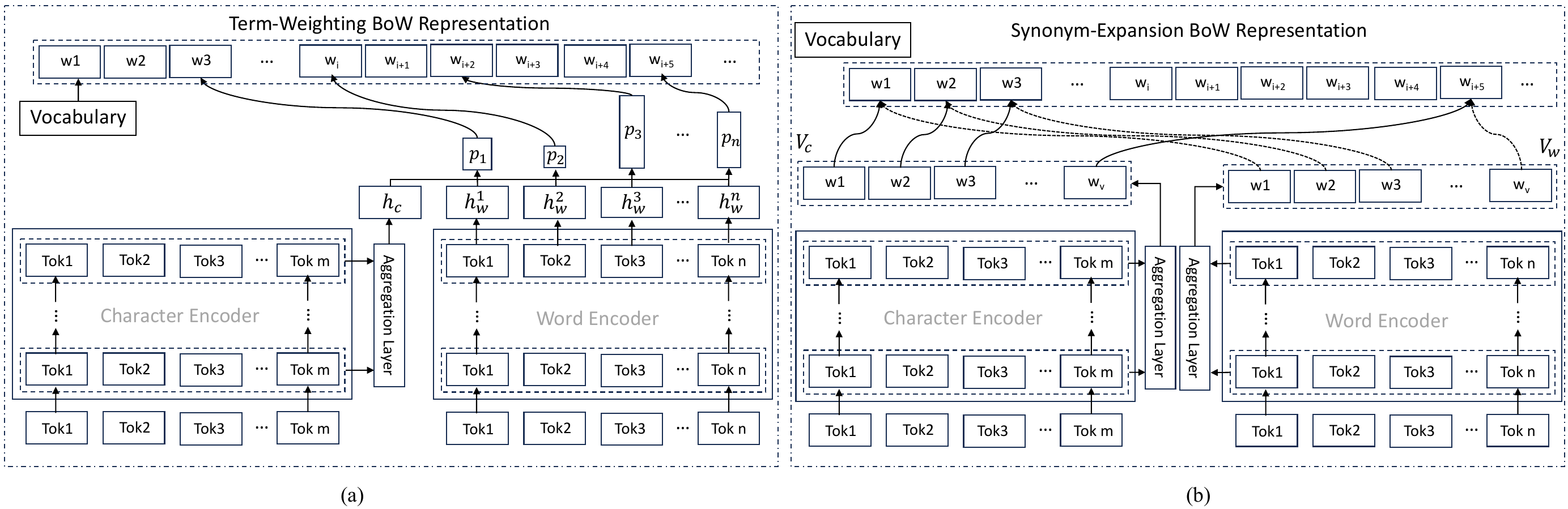}
\caption{ An overview of the DeepBoW model. Figure (a) shows the architecture that encodes the input text into the Term-Weighting BoW representation, which gathers the attention weight of each word as its weight in the term-weighting BoW representation. Figure (b) shows the architecture that encodes the input text into the Synonym-Expansion BoW representation, which generates sparse BoW representation from character embedding and word embedding respectively, and aggregates these two representations as the synonym-expansion BoW representation.}
\label{overview}
\end{figure*}
\section{Methodology}
\label{sec:method}
\subsection{Overview}

The proposed DeepBoW model is based on the two-tower architecture, which encodes the query and document separately and computes the semantic relevance score with the representations of query and document. Different from other text relevance models with dense embeddings, our model encodes the query and document into the Bag-of-Words vectors and calculates the relevance score from sparse BoW representations. 

In this section, we introduce the components of our model in detail. We first describe the multi-granularity encoder to aggregate the character-grained feature and word-grained feature. Next, we introduce two different sparse BoW representations including the term-weighting BoW and the synonym-weighting BoW. Then, we show how to use N-gram hashing to reduce the semantic loss from word segmentation and enhance the quality of the sparse BoW representation. Finally, we describe the training process of our model in detail and show the deployment of our DeepBoW model to the online e-commerce search system. Figure \ref{overview} shows an overview of the DeepBoW model.

\subsection{Multi-Granularity Encoder}

The text encoder aims to obtain the input text's contextual representations.
We choose Transformer encoder \cite{NIPS2017_transformer} as our sentence encoder because of its excellent performance in many tasks. The Transformer encoder is a stack of $L$ identical layers, and each layer includes a multi-head self-attention and a fully connected feed-forward network. For the input senquence $S$, we obtain the output encoding matrix of $i$-th layer as $H^i = \{h^i_1, h^i_2, \cdots, h^i_l \}$, where $h^i_j \in \mathbb{R}^d$  is the word embedding vector and $l$ is the number of words in $S$. Same to ReprBERT \cite{10.1145/3534678.3539090}, we aggregate the output of each layer as the text encoding representation according to:

\begin{equation}
    \begin{aligned}
    \tilde{h}^i &= \left(\frac{1}{n}\sum_{h^i_j\in H_i} h^i_j\right) \mathbf{W_m} + \mathbf{b_m} \\
    h &= \left[\bigparallel_{i=1}^{L} h^i \right] \mathbf{W_{agg}}  + \mathbf{b_{agg}}
    \end{aligned}
\end{equation}

\noindent where $\mathbf{W_m} \in \mathbb{R}^{d \times d}, \mathbf{W_{agg}} \in \mathbb{R}^{L\cdot d \times d}, \mathbf{b_m}, \mathbf{b_{agg}} \in \mathbb{R}^d$, $\bigparallel$ is the concatenate operation.

 Usually the character-based model performs better than the word-based model in Chinese NLP tasks \cite{li-etal-2019-word-segmentation}, and most Chinese-BERT models are character-based Transformer architecture. However, since our model intends to catch the relationship between the semantic and the word of the sentence, we propose to encode not only the character-segmentation sequence but also the word-segmentation sequence separately.
For convenience, we denote the character-segmentation sequence and the word-segmentation sequence of the input text as $S_c = \{c_1, c_2, \cdots, c_m\}$ and $S_w = \{w_1, w_2, \cdots, w_n\}$, respectively, where $c_i$ and $w_i$ are the indices of the token in the vocabulary. The output encoding matrices of the character-segmentation sequence and the word-segmentation sequence are $H_c = \{h_c^1, h_c^2, \cdots, h_c^m\}$ and $H_w = \{h_w^1, h_w^2, \cdots, h_w^n\}$, where $m$ and $n$ are the lengths of the character-segmentation and the word-segmentation sequence. The text encoding representation of these two sequences are $h_c$ and $h_w$.

\subsection{Sparse BoW Representation}

Unlike traditional two-tower architecture models that encode text into the "embedding" which is a dense distributed representation, our model encodes the text into the sparse BoW representation. The sparse BoW representation is a set of word-weight pairs, where each word corresponds to a weight that indicates the importance or the relevance of this word to the input text. In this section, we introduce two different sparse BoW representations: term-weighting BoW representation and synonym-weighting BoW representation, and describe the module to generate these two sparse BoW representations in detail.

\subsubsection{Term-Weighting BoW Representation}
\
\newline
\begin{CJK*}{UTF8}{gbsn}
In the e-commerce search system, the query inputted by user may contain some redundant or unrelated words. These words can be excised with negligible impact on the text semantic. For example, for the input query from Taobao like "2024年 夏季 适合 准妈妈 孕妇 套装"\footnote{This query means ``2024 Summer Pregnant Women's Clothing''.}, "准妈妈" and "孕妇" both mean a pregnant woman, but "孕妇" is more accurate than "准妈妈" at semantic level as the latter word is polysemous and more colloquial. "适合" which means suitable, can be regarded as a stop word in the e-commerce scenario. So "准妈妈" and "适合" can be discarded and the other words should be retained.
\end{CJK*}

Term-Weighting BoW includes all words of the input text, and each word is assigned a weight that indicates its significance within the text's semantics. Key words like brand and category should have greater importance weights than the other words. Figure \ref{overview} (a) shows the architecture to generate the term-weighting BoW representation. Then, the term-weighting BoW representation can be produced as follows:

\begin{equation}
\begin{aligned}
p_i &= \frac{\exp \left(h_c^{\top} h_w^i\right)}{\sum_{h_w^j\in H_w} \exp \left(h_c^{\top} h_w^j\right)} \\
\operatorname{BoW_{tw}}(S_w) &:= \{w_i: p_i, w_i \in S_w\}
\end{aligned}
\end{equation}

\noindent We define the $\operatorname{BoW_{tw}}(\cdot)$ as the term-weighting BoW representation of $\cdot$. $p_i$ is the importance weight of $w_i$ in $S_w$, and $\sum_i p_i = 1$.

\subsubsection{synonym-expansion BoW Representation}
\
\newline
    Since queries in e-commerce search systems are entered by lay users, they may differ from the product descriptions and include colloquialisms and polysemes. Some adjectives or category words may also have synonyms. Term-weighting BoW representation can only compute the importance weight of each word in the text, but is unable to add relevant words and synonyms. Synonym expansion can greatly improve the performance of the e-commerce search system. Therefore, we propose the synonym-expansion BoW representation to enhance the retrieval performance of the sparse BoW representation. Figure \ref{overview} (b) shows the architecture to generate the synonym-expansion BoW representation.  

We sample $v$ words from the training corpora as the vocabulary $\mathbb{V}$ according to the frequency of the word. Our model leverages the relevance between these words and the input text to represent the semantics of the query and the products. First, our model aggregates the word-based text encoding representation and the character-based text encoding representation as follows:

\begin{equation}
\begin{aligned}
\tilde{h}_w &= \sum_{i=1}^n p_i\cdot h_w^i \\
V_c &= \sigma \left(h_c\mathbf{W_{c}} + \mathbf{b_{c}}\right) \\
V_w &= \sigma \left(\left[h_c||\tilde{h}_w\right]\mathbf{W_{w}} + \mathbf{b_{w}} \right) \\
\label{projection}
\end{aligned}
\end{equation}

\noindent where $\mathbf{W_c} \in \mathbb{R}^{d \times v},\mathbf{W_w} \in \mathbb{R}^{2d \times v}$ and $\mathbf{b_c},\mathbf{b_w} \in \mathbb{R}^{v}$. $\sigma$ is the sigmoid function. Then, the synonym-expansion BoW Representation of the input text is as follows:

\begin{equation}
\begin{aligned}
p_g &= \sigma \left( \left[h_c||\tilde{h}_w\right]\mathbf{W_{g}} + \mathbf{b_{g}} \right) \\
\operatorname{BoW_{SE}}(S_w) &:= \{t: V_c(t), t \in \mathbb{V} - S_w \} \cup \\
    &\{t: p_gV_c(t) + (1 - p_g)V_w(t), t \in S_w \cap \mathbb{V}\}
\end{aligned}
\end{equation}

\noindent where $V(i)$ denotes the $i$-th value of $V$. We define the$ \operatorname{BoW_{SE}}(\cdot)$ as the synonym-expansion BoW representation of $\cdot$. $t$ is the word (actually is the index of the word) in $\mathbb{V}$. the corresponding weight is in $[0,1]$, which can be regarded as the relevance score between this word and the input text.

\subsection{N-gram Hashing Vocabulary}

In the preceding section, we describe the sparse BoW representation in detail. Unfortunately, due to the limitation of model's parameter size,  we can only leverage the vocabulary within a limited number of words. Using the ``\verb|[UNK]|'' to replace all Out-Of-Vocabulary (OOV) words may lead to significant semantic loss. To mitigate this issue, we introduce an ensemble of hashing tokens into the vocabulary, where the OOV word can be replaced with its hashing tokens\footnote{For the word $w$, we leverage $\operatorname{MD5}(w)\%B$ as its hashing token, where $B$ is the hashing bucket number.}. 

Semantic loss may occur between the raw text and its BoW representation, particularly when syntactically cohesive phrases are fragmented during the word segmentation process. This issue can lead to misalignment for the essential semantics such as product types, entity names, or brand identifiers in query/product. For example, the brand name ``\verb|L'ORÉAL Paris|'' could be inaccurately divided into separate tokens during word segmentation. To address this problem, we introduce an N-gram hashing vocabulary strategy. Concretely, N-gram phrases are incorporated into the text's BoW and are subsequently replaced with their respective hashing tokens, analogous to the treatment of OOV words. The significance of a particular N-gram phrase is directly proportional to the frequency of its occurrence within relevant query-product pairs in the corpora. Our model is equipped to ascertain the importance of these N-gram hashing tokens through the analysis of large-scale corpora. Consequently, the semantics of these N-gram phrases are retained within the sparse BoW representation.

\subsection{DeepBoW Relevance Model}

In this section, we describe the method to compute the relevance score between the query and the product from the sparse BoW representations. Note that in the search engine scenario the product should match all the semantics of query, while conversely the query does not need to match all the semantics of the product. Accordingly, we encode the query as the term-weighting BoW representation while encode the product as the synonym-expansion BoW representation. The relevance score of the query/product can be calculated as follows:

\begin{equation}
R_t(Q,D) = \sum_{(w:p) \in \operatorname{BoW_{TW}}(Q) \atop (t:g) \in \operatorname{BoW_{SE}}(D)} \sum_{w=t} p \times g 
\label{r_t}
\end{equation}

\noindent where $Q$ is the query, $D$ is the product, and $R_t(Q,D)$ is the relevance score between $Q$ and $D$. We call our DeepBoW model with this relevance score as \textbf{DeepBoW(Q-Weight)}.

We leverage the cross-entropy loss between the output score and the ground truth to train our model. In addition, we also optimize the L2 norm of $\operatorname{BoW_{SE}}(D)$ to enhance the sparsity of the synonym-expansion BoW representation, so that only the most relevant words can get high scores. The loss function is as follows:

\begin{equation}
\begin{aligned}
norm &= \left(\sum_{(t:g) \in \operatorname{BoW_{SE}}(D)} g^2 \right)^{\frac{1}{2}} \\
loss_t &= \operatorname{CE}\left(R_t, label\right) + \frac{1}{v} norm
\end{aligned}
\end{equation}

\noindent where $\operatorname{CE}$ is the cross-entropy loss, $label$ is the ground truth of the training data.
We can also encode the query as the synonym-expansion BoW representation to improve the performance of recall. The relevance score is as follows:
\begin{equation}
\begin{aligned}
C &= \sum_{(w:p) \in \operatorname{BoW_{SE}}(Q)} p \\
R_s(Q,D) &= \frac{1}{C}\sum_{(w:p) \in \operatorname{BoW_{SE}}(Q) \atop (t:g) \in \operatorname{BoW_{SE}}(D)} \sum_{w=t} p \times g 
\end{aligned}
\label{r_s}
\end{equation}
\noindent We call our DeepBoW model with this relevance score as \textbf{DeepBoW(Q -Synonym)}.
Different from $loss_t$, we also leverage the bag-of-words of the query to train the product's synonym-expansion BoW representation. The loss should be modified as follows:

\begin{equation}
\begin{aligned}
\operatorname{BoW_{avg}}(S_w) &:= \{w_i: 1/n, w_i \in S_w\} \\
R_{avg}(Q,D) &= \sum_{(w:p) \in \operatorname{BoW_{avg}}(Q) \atop (t:g) \in \operatorname{BoW_{SE}}(D)} \sum_{w=t} p \times g\\
loss_s = \operatorname{CE}(R_s, &label) + \operatorname{CE}\left(R_{avg}, label\right) + \frac{1}{v} norm
\end{aligned}
\end{equation}

For both $loss_t$ and $loss_s$, we optimize the difference between the query's sparse BoW representation and the product's sparse BoW representation. This can align the vocabulary between the query and the product.

 \subsection{Online Deployment}

Since most online search systems have strict latency limitations, we pre-compute the sparse BoW representation of the queries and the products offline. We discard the word-weight pairs in the sparse BoW representation whose weights are lower than the given threshold to optimize memory usage. We sort the word-weight pairs offline by the word index. The relevance score of two sparse BoW representations can be calculated by using the two-pointer algorithm. Then the time complexity of Eq.\ref{r_t} and Eq.\ref{r_s} can be optimized to $\mathcal{O}(N)$, which is much faster than the state-of-the-art deep relevance model \cite{10.1145/3534678.3539090}. Although some deep relevance models with cosine similarity can achieve comparable efficiency, the performance of these models is much lower than our model as shown in the section \ref{sec:result}.

\section{EXPERIMENTS}
\label{sec:experiments}
\subsection{Dataset}

There is no public dataset and benchmark for the Chinese e-commerce relevance task, so we conduct experiments on three different types of industrial datasets to learn the DeepBoW model. The first is a large-scale \textbf{Human-Annotation} dataset
which contains query-product pairs sampled from the Taobao search logs. Each query-product pair is labeled as Good (relevant) or Bad (irrelevant) by experienced human annotators. This is a daily task running in Taobao, which has accumulated more than 8 million labeled samples. We split the human-annotated datasets into training, validation and test sets, as detailed in Table~\ref{statistic}. 

\begin{table}[htb]
\centering
\scalebox{0.95}{
\begin{tabular}{c|ccccc}
\toprule[1.2pt]

\textbf{Dataset} & \textbf{sample} &\textbf{query} & \textbf{product} & \textbf{Good} & \textbf{Bad} \cr
\hline
Train & 7,439,823 & 463,387 & 6,068,089&6,092,745&1,347,078\cr
Valid & 372,981   & 98,254	& 202,192  &305,447	 &67,534 \cr
Test  & 984,175	  & 134,564	& 883,691  &745,524	 &238,651\cr
\bottomrule[1.2pt]
\end{tabular}}
\caption{Statistic for human-annotation dataset.}
\label{statistic}
\end{table}

The second dataset for training is built by knowledge distillation, similar to \citet{10.1145/3534678.3539090}. We leverage the training set of the human-annotation dataset to finetune the StructBERT \cite{Wang2020StructBERT:} model, which results in an interaction-based teacher model with strong performance. Then the teacher model predicts the relevance scores of the large unlabeled query-product pairs sampled from the search logs of Taobao within a year. This training dataset is denoted as ``\textbf{Search-Logs}'' in Table \ref{result}.
Third, we also sample click-through data from search logs and investigate the performance of our model on this training set. We denote this dataset as ``\textbf{Click-Through}'' in Table \ref{result}.
Although these are training datasets of different sources, we all use the human-annotation validation and test dataset to evaluate the model performance.

\subsection{Training Details}

We employ Transformer as both the character-based encoder and the word-based encoder. We reduce the total 12-layer encoder to improve efficiency. After balancing the effectiveness and efficiency, our model adopts 2 layers that can still achieve competitive performance. 
We select the top 50000 words as the vocabulary $\mathbb{V}$ according to the word's frequency in corpora, and we also add another 10000 hashing tokens into the vocabulary for the OOV words and the N-gram phrases.

We use ``\verb|PyTorch|'' to implement our model and train the model with Adam optimizer. The hyper-parameters of Adam optimizer are $\beta_1 = 0.9,\beta_2 = 0.999,\epsilon = 10^{-8}$ and the learning rate is set to 0.0001. Query-document pairs are batched together by approximate sequence length. Each training batch contains a set of sentence pairs with about 50000 tokens. The hyper-parameters and the best model are chosen from the experimental results on the validation set. We train our model on 2 Tesla V100 GPU and it usually takes about 3 days for the model to converge. The convergence is reached when the ROC-AUC does not improve on the validation set. 

\begin{table*}
\center
\scalebox{1}{
\begin{tabular}{l|cc|cc|cc}
\toprule[1.2pt]

\textbf{Model} & \multicolumn{2}{c|}{\textbf{Human-Annotation}} & \multicolumn{2}{c|}{\textbf{Search-Logs}} & \multicolumn{2}{c}{\textbf{Click-Through}} \cr
& ROC-AUC & Neg PR-AUC & ROC-AUC & Neg PR-AUC& ROC-AUC & Neg PR-AUC \cr
\hline
\textbf{Interaction-Based Models:} & & & & & & \cr
BERT\cite{devlin-etal-2019-bert} & 0.850 & 0.662 & - & - & - & - \cr
RoBERTa\cite{Liu2019RoBERTaAR} & 0.906 & 0.692 & - & - & - & -  \cr
StructBERT\cite{Wang2020StructBERT:} & \textbf{0.923} & \textbf{0.721} & - & - & - & - \cr
\hline
\textbf{Two-Tower Models:} & & & & & & \cr
Siamese BERT & 0.765 & 0.565 & 0.821 & 0.648 & - & - \cr
MASM\cite{10.1145/3442381.3450129} & 0.795 & 0.484 & 0.793 & 0.582 & 0.615& 0.283 \cr
Poly-Encoder\cite{Humeau2019PolyencodersAA} & 0.808 & 0.623 & 0.846 & 0.605 & - & - \cr
DSSM\scalebox{0.75}{RoBERTa} & 0.873 & 0.673 & - & - & - & -  \cr
DSSM\scalebox{0.75}{StructBERT} & 0.858 & 0.658 & - & - & - & -  \cr
ReprBERT \cite{10.1145/3534678.3539090}& 0.832 & 0.543 & 0.894 & 0.702 & 0.798 & 0.521 \cr
ReprBERT \scalebox{0.75}{+Cosine Similarity} & 0.798 & 0.452 & 0.847 & 0.601 & 0.727 & 0.399 \cr

\hline
\textbf{Ours:} & & & & & & \cr

DeepBoW(Q-Weight) & 0.874 & 0.665 & 0.908 & 0.705 & 0.803 & 0.579 \cr
DeepBoW(Q-Weight) \scalebox{0.75}{+128-Trunc} & 0.865 & 0.645 & 0.899 & 0.698& 0.787 & 0.566 \cr
DeepBoW(Q-Weight) \scalebox{0.75}{+0.4-Trunc} & 0.869 & 0.658& 0.906 & 0.701& 0.796 & 0.572 \cr
DeepBoW(Q-Synonym) & \underline{0.880} & \underline{0.674} & \textbf{0.914} & \textbf{0.712} & \textbf{0.812} & \textbf{0.585}\cr
DeepBoW(Q-Synonym) \scalebox{0.75}{+128-Trunc} & 0.873 & 0.670 & 0.906 & 0.705 & 0.799 & 0.571\cr
DeepBoW(Q-Synonym) \scalebox{0.75}{+0.4-Trunc} & 0.877 & 0.672& 0.911 & 0.710 & 0.807 & 0.575 \cr

\bottomrule[1.2pt]
\end{tabular}}
\caption{\label{result} Comparison results on test set. Best scores are in bold. \scalebox{0.75}{+128-Trunc} means keeping 128 largest terms according to the value of the word-weight pair. \scalebox{0.75}{+0.4-Trunc} means discarding the terms that the value is smaller than 0.4. We only finetune the pre-trained based models on the Human-Annotation dataset and do not evaluate these models in the other two training sets, since we leverage StructBERT as teacher model to label the search-logs dataset. We do not evaluate the performance of MASM and Poly-Encoder on Click-Through training dataset, because the two models do not converge on the other two training datasets.}
\end{table*}

\subsection{Baseline}

We explore the performance of DeepBoW(Q-Weight) and DeepBoW (Q-Synonym) respectively. The main difference between the two methods is for DeepBoW(Q-Synonym) we leverage the synonym-expansion BoW representation to replace the term-weighting BoW representation for the query. To reduce memory usage and computation, we truncate the BoW representation to make it sparse. There are two methods to truncate the sparse BoW representation, one is to keep the $k$ largest words according to their respective values, and the other is to discard the terms whose values are smaller than the giving threshold.

In addition, we adopt several state-of-the-art methods for comparison. BERT \cite{devlin-etal-2019-bert}, RoBERTa \cite{Liu2019RoBERTaAR} and StructBERT \cite{Wang2020StructBERT:} belong to the interaction-based architecture which is also known as the cross-encoder architecture. Siamese BERT \cite{10.1145/3308558.3313707}, MASM \cite{10.1145/3442381.3450129}, Poly-encoders \cite{Humeau2019PolyencodersAA} and ReprBERT \cite{10.1145/3534678.3539090} belong to the two-tower architecture which is also known as the bi-encoder architecture. Besides, we investigate the performance of ReprBERT with cosine similarity score instead of MLP for online computation of relevance from query/product embeddings. For fair comparison, we also leverage the pre-trained model like RoBERTa and StructBERT as the encoder of the two-tower model. These models are also baselines in our experiment, which denote as DSSM\scalebox{0.75}{RoBERTa} and DSSM\scalebox{0.75}{StructBERT}.

\subsection{Evaluation Metrics}

We evaluate our model on both offline and online metrics. In offline evaluation, since the human annotation is binary, the task is evaluated as a classification task. The Receiver Operator Characteristic Area Under Curve (ROC-AUC) is widely adopted in text relevance tasks \citep{10.1145/3326937.3341259,10.5555/3060832.3061030,10.1145/1143844.1143874}. Note that in the e-commerce relevance scenario, most instances are positive and we are more concerned about negative instances. Therefore the PR-AUC used in this paper is the negative PR-AUC that treats Bad as 1 and Good as 0 following \citet{10.1145/3442381.3450129,10.1145/3534678.3539090}. This metric is denoted as Neg PR-AUC. 

Besides, we also evaluate the different model complexity of parameters and online computation efficiency. The FLOPs / token is computed according to \citet{molchanov2017pruning} which shows the floating-point operations per second (FLOPs) when there is only 1 token being considered. The plus sign separates the online and offline calculation FLOPs, which means the former part of computation can be pre-computed offline. Memory indicates the online memory overhead for storing pre-computed query and product vectors where we use vector size for comparison. In online evaluation, we use the rate of Good annotated by human annotators and the number of transactions as the evaluation metrics. The query-product pairs for human relevance judgment are randomly sampled from the online search logs according to the amount of Page View (PV) as the sample weight.

\subsection{Results}
\label{sec:result}
Table \ref{result} presents an evaluative comparison across various models. Our DeepBoW model demonstrates robust performance across three different training sets. For human-annotation dataset, StructBERT has the best performance, and interaction-based models outperform two-tower models. Unfortunately, it is infeasible to deploy the interaction-based model in industrial system because of prohibitive computation and resource requirements. The ROC-AUC and Neg PR-AUC of two-tower models are much lower than the pre-trained model, because the human-annotation data is insufficient and the pre-trained model can introduce extra knowledge. Nonetheless, our DeepBoW model still outperforms other two-tower models.

The data enhancement sampled from the search logs and labeled by the teacher model can greatly improve the performance of two-tower models. Our model achieves the best performance around the two-tower models in search-logs training sets. Click-through data is used to train the relevance model in some cases where lack of human-annotation training data. The models trained on the click-through data get weak performance. The main reason is that the click-through data in e-commerce is much more noisy and misleading, which is not only affected by the query-product relevance but also by many factors including price, attractive titles, or images. Even so, our model also performs better than the other models since it explicitly encodes the semantics to the sparse bag-of-words while the other models may capture the personalized information beyond the textual feature.

Our proposed architecture can truncate the sparse BoW representation to reduce memory usage in the online search system. We can either truncate the sparse BoW representation to the fixed length or discard the word-weight pairs whose values are smaller than a given threshold. Benefiting from the interpretability of our sparse BoW representation, both truncation methods achieve competitive performance and only produce a slight loss of performance. We further explore the distribution of the sparse BoW representation as case study in Appx. \ref{appx_case}.

\begin{table}
\centering
\scalebox{0.99}{
\begin{tabular}{l|ccc}
\toprule[1.2pt]

\textbf{Model} & \textbf{Params} &\textbf{FLOPs / Token} & \textbf{Mem} \cr
\hline
BERT & 101.2M & 182M & 0 \cr
RoBERTa & 101.2M & 182M & 0 \cr
StructBERT & 101.2M & 182M & 0  \cr
Siamese BER & 101.2M & 91M + 1.5K & 768 \cr
MASM & 76.8M & 674K & 640 \cr
Poly-Encoder & 101.2M & 182M + 97.5K & 768 \cr
ReprBERT & 30.6M & \textbf{30.4M} + 296K & 256 \cr
\hline
\textbf{Ours:} & & & \cr

DeepBoW \scalebox{0.75}{+128-Trunc} & \scalebox{0.9}{33.4M + 126M} & 159.4M + 128 & 128\cr
DeepBoW \scalebox{0.75}{+0.4-Trunc} & \scalebox{0.9}{33.4M + 126M} & 159.4M + \textbf{28} & 28/144\cr

\bottomrule[1.2pt]
\end{tabular}}
\caption{\label{efficiency} The models' efficiency. The Params of our models include the encoder and the final vocabulary projection ($\mathbf{W_c}$ and $\mathbf{W_w}$ in Eq.\ref{projection}). The Mem of DeepBoW \scalebox{0.75}{+0.4-Trunc} is the mean value throughout the whole corpora (query/product).}
\end{table}

Table \ref{efficiency} shows the parameter, computation and memory consumption of each model. While our model has a considerable number of parameters since we project the dense vector into the vocabulary (78.8\% parameters about 126M come from $\mathbf{W_c}$ and $\mathbf{W_w}$ in Eq.\ref{projection}), DeepBoW is the most efficient at online inference. Our model only uses CPU to compute the relevance score while the other models need GPU to speed the inference. Table \ref{Time} shows the inference time of each model. The experiments are performed on a local CPU platform. We report the average inference time of the model to score 1000 products per query. The results show that our model is much faster than the ReprBERT \cite{10.1145/3534678.3539090} which has been deployed in the Taobao search system.

\subsection{Ablation Study}

We perform the ablation study on the human-annotation dataset to investigate the influence of different modules in our model. We investigate the performance of the 2-layer encoder and the 6-layer encoder. Besides, we remove the word encoder and the character encoder separately to show the importance of encoding at both word-level and character-level. We also employ different capacities of vocabulary and hashing tokens in the sparse BoW representation. To further explore the sparsity of BoW representation, we remove the $l_2$ norm from the loss function. This may lead to a sparsity deterioration in the sparse BoW representation. 

Table \ref{ablation_study} shows the results of the ablation study. 
We can see that each module in our model does contribute to the overall performance. The performance of our model has a significant deterioration if either the word encoder or the character encoder is removed. Increasing the capacity of vocabulary or hashing tokens cannot improve the model's performance, because it may lead to insufficient training for each token. Optimizing our model without the $l_2$ norm loss can lead to slight performance decline. There is almost no difference between the 2-layer and 6-layer encoders in our model.

\begin{table}
\centering
\scalebox{1}{
\begin{tabular}{l|c}
\toprule[1.2pt]
\textbf{Model} & \textbf{Inference Time (ms)} \cr
\hline
StructBERT & 321,408 \cr
ReprBERT   & 29,164 \cr
DeepBoW  & \textbf{0.732} \cr

\bottomrule[1.2pt]
\end{tabular}}
\caption{\label{Time} Inference time of different models that score 1000 query-product pairs.}
\end{table}

\begin{table}
\centering
\scalebox{1}{
\begin{tabular}{l|cc}
\toprule[1.2pt]

\textbf{DeepBoW} & \textbf{ROC-AUC} &\textbf{Neg PR-AUC} \cr
\hline
w/ 2 layers & 0.914 & 0.712 \cr
w/ 6 layers & 0.911 & 0.717 \cr
w/ 10w vocab & 0.907 & 0.701 \cr
w/ 5w hashing tokens & 0.899 & 0.693 \cr
w/o $l_2$ norm loss & 0.886 & 0.685\cr
w/o word encoder & 0.847& 0.611 \cr
w/o character encoder & 0.821& 0.583 \cr

\bottomrule[1.2pt]
\end{tabular}}
\caption{\label{ablation_study} Ablation study of components of DeepBoW}
\end{table}

\subsection{Online Evaluation}

Online A/B testing is also conducted to evaluate our DeepBoW model, by replacing the online ReprBERT model with the DeepBoW model for comparison. Both experiments take about 2.5\% proportion of Taobao search traffic, and the A/B testing lasts for a month. As a result, DeepBoW improves the number of transactions by about 0.4\% on average. The daily human annotation results show that DeepBoW also improves the rate of relevance by 0.5\%. Online A/B testing verifies the proposed DeepBoW is superior to previous state-of-the-art models, and can achieve significant online profit considering the extremely large traffic of Taobao every day.

Our DeepBoW model has already served the entire Taobao search traffic. After pre-computing the representations of queries and products, the online serving latency can be optimized to as low as 4ms on the distributed computing system with CPUs. This is much faster than the previous online relevance serving model ReprBERT \cite{10.1145/3534678.3539090} of 10ms with GPUs and can satisfy the extremely large traffic of Taobao.

\section{Conclusion and Future Work}

In this paper, we study an industrial task of measuring the semantic relevance of queries and products. We propose the DeepBoW relevance model, which is an efficient and interpretable relevance architecture for Chinese e-commerce search system. Our model encodes the query and product as a set of word-weight pairs, which is called the sparse BoW representation. The model is evaluted on three different training datasets, and the results show that our model achieves promising performance and efficiency. The model has been deployed in the Taobao search system.

In future work, we will explore integrating external knowledge into the DeepBoW relevance model to improve the performance. The proposed model can also be evaluated on datasets of other language.
\bibliographystyle{ACM-Reference-Format}
\balance
\bibliography{kdd24}

\appendix
\begin{table*}
    \centering
    \scalebox{0.9}{
    \begin{CJK*}{UTF8}{gbsn}
    \begin{tabular}{p{18cm}}
    \hline
    \multicolumn{1}{c}{\textbf{Cases of the sparse BoW representation}} \cr
    \hline
    \cr
    \textbf{Query:} 小香风连衣裙 \cr
    \boldmath $\operatorname{BoW_{SE}(Query)}$: \unboldmath [(`连衣裙', 0.30148), (`高级感', 0.25785), (`小香风', 0.2277), (`新款', 0.21297)] \cr\cr
    \textbf{Product:} 高级感秋冬装小香风长袖V领针织连衣裙通勤复古赫本小黑裙打底裙 连衣裙 18-24周岁 针织布 纯色 通勤 V领 高腰 套头 黑色 酒红色 粉红色 常规 单件 复古 A字裙 秋冬 2023年秋季 长袖 中裙 链条 \cr
    \boldmath $\operatorname{BoW_{SE}(Product)}$: \unboldmath [(`品质', 1.0), (`v领', 1.0), (`秋', 0.99999), (`高腰', 0.99999), (`高级感', 0.99999), (`通勤', 0.99999), (`秋季', 0.99998), (`秋冬', 0.99998), (`秋冬装', 0.99998), (`黑色', 0.99995), (`赫本', 0.99993), (`单件', 0.99993), (`18', 0.99992), (`中裙', 0.99991), (`针织布', 0.9999), (`长袖', 0.99985), (`打底', 0.99981), (`复古', 0.9998), (`黑裙', 0.9998), (`小黑裙', 0.99977), (`针织', 0.99972), (`纯色', 0.9997), (`常规', 0.99968), (`a字裙', 0.99961), (`链条', 0.99957), (`24', 0.99954), (`字', 0.99944), (`粉红', 0.9994), (`新款', 0.99934), (`粉红色', 0.9992), (`时尚', 0.99897), (`气质', 0.99862), (`打底裙', 0.99824), (`秋款', 0.9981), (`女士', 0.99805), (`女裙', 0.99726), (`2023', 0.99721), (`早秋', 0.99687), (`小香风', 0.99657), (`春秋', 0.99627), (`2023年', 0.99622), (`女装', 0.99621), (`套头', 0.99618), (`酒红色', 0.99616), (`黑色连衣裙', 0.99583), (`针织连衣裙', 0.99577), (`新款连衣裙', 0.99536), (`装', 0.99512), (`冬装', 0.99502), (`秋装', 0.9946), (`连身', 0.9943), (`秋冬连衣裙', 0.99379), (`针织裙', 0.99247), (`气质连衣裙', 0.99237), (`18-24', 0.99174), (`酒红', 0.99117), (`女', 0.99094), (`袖子', 0.99014), (`自动充气垫', 0.98937), (`轻熟', 0.98916), (`2019', 0.98725), (`韩版', 0.98703), (`春季', 0.98666), (`布', 0.98649), (`内搭', 0.98606), (`高腰连衣裙', 0.98587), (`黑', 0.98563), (`a字', 0.98546), (`修身', 0.98471), (`连衣裙', 0.98461), (`裙子', 0.98451), (`裙', 0.98445), (`初秋', 0.98441), (`休闲连衣裙', 0.98398), (`流行', 0.98365), (`红色连衣裙', 0.98294), (`宽松连衣裙', 0.9822), (`衣服', 0.9794), (`显瘦', 0.97937), (`秋冬新款', 0.97758), (`冬季', 0.97752), (`女连衣裙', 0.97674), (`个子', 0.97423), (`红色裙子', 0.9712), (`名媛', 0.96755), (`名媛连衣裙', 0.96739), (`女生', 0.96676), (`打底衫', 0.96633), (`连身裙', 0.9661), (`洋气', 0.96411), (`名媛气质连衣裙', 0.96329), (`修身连衣裙', 0.96265), (`衣', 0.96137), (`打底连衣裙', 0.95868), (`冬款', 0.95856), (`毛衣裙', 0.95822), (`长袖连衣裙', 0.95057), (`连衣裙子', 0.95046), (`搭', 0.94854), (`秋冬裙子', 0.94758), (`休闲', 0.94602), (`设计感', 0.94598)] \cr\cr
    \textbf{Relevance Score:} \quad $0.30148 \times 0.98461 + 0.25785 \times 0.9999 + 0.2277 \times 0.99657 + 0.21297 \times 0.99934 = 0.9944$ \cr
    \cr
    \hline
    \cr
    \textbf{Query:} 秋冬床上四件套 \cr
    \boldmath $\operatorname{BoW_{SE}(Query)}$: \unboldmath [(`四件', '0.343'), (`床上', '0.13778'), (`套', '0.06248'), (`秋冬', '0.09616'), (`床上四件套', '0.10872'), (`四件套', '0.202')] \cr\cr
    \textbf{Product:} 秋冬季加厚牛奶绒床上四件套珊瑚绒被套双面法兰绒加绒床单三件套 卡丝迪尔家纺旗舰店 床品套件/四件套/多件套 Kiss Dear/卡丝迪尔 蓄热保暖 牛奶绒 卡通动漫 卡通 中国大陆 2021年秋季 床单式 床笠式 大众\cr
    \boldmath $\operatorname{BoW_{SE}(Product)}$: \unboldmath [(`珊瑚绒被套', 1.0), (`珊瑚绒', 1.0), (`江苏', 1.0), (`2021年', 1.0), (`大陆', 0.99999), (`多件套', 0.99999), (`蓄热', 0.99999), (`床笠式', 0.99998), (`冬季', 0.99997), (`江苏省', 0.99997), (`秋冬季', 0.99993),  (`床品', 0.99988), (`三件套', 0.99979), (`保暖', 0.99979), (`床上', 0.99979), (`加厚', 0.99975), (`被套', 0.99975), (`四件', 0.99965), (`南通', 0.99955), (`床品套件', 0.99948), (`套', 0.99926), (`件套', 0.99919), (`牛奶', 0.99906), (`秋', 0.99891), (`床单三件套', 0.9984), (`卡通', 0.99832), (`动漫', 0.9981), (`dear', 0.99762), (`床上三件套', 0.99754), (`双面', 0.99744), (`床上四件套', 0.99703), (`加绒', 0.99627), (`套件', 0.99615), (`kiss', 0.99586), (`床单', 0.9952), (`珊瑚', 0.99498), (`自动充气垫', 0.9935), (`床单四件套', 0.98934), (`被罩', 0.98758), (`床笠', 0.98725), (`2021', 0.98496), (`床盖', 0.978), (`双面绒', 0.96437), (`毛绒', 0.95704), (`秋季', 0.94864), (`冬天', 0.93963), (`毛毛', 0.9209), (`秋冬', 0.90725), (`用品', 0.89087), (`四件套', 0.88608), (`儿童四件套', 0.87247), (`珊瑚绒床单', 0.86876), (`冬', 0.85713), (`冬款', 0.85588), (`床品四件套', 0.82923), (`单人床', 0.82689), (`绒面', 0.82627), (`单人', 0.81626), (`潮牌', 0.78414), (`绒', 0.77402), (`宿舍', 0.76135), (`少女', 0.6812), (`简约', 0.60279), (`被子', 0.59161), (`家纺', 0.56128), (`秋冬款', 0.55233), (`絲', 0.54342), (`旗舰店', 0.54124), (`冬装', 0.52361), (`厚', 0.51875), (`高档', 0.51707), (`单件', 0.51121), (`薄绒', 0.50859), (`卡通四件套', 0.48766), (`床上用品', 0.48612), (`枕套', 0.46126), (`双人', 0.45578), (`4件套', 0.44235), (`套装', 0.43786), (`保暖衣', 0.42136), (`全套', 0.40858), (`睡', 0.39326), (`珊瑚绒睡衣', 0.38609), (`床套', 0.35582), (`丝绒', 0.34883), (`加绒裤', 0.32625), (`1.5', 0.30884), (`学生', 0.29069), (`神器', 0.28852), (`红色系', 0.28552), (`网红', 0.28548), (`毛茸茸', 0.28526), (`必备', 0.28513), (`装饰品', 0.28248), (`冬季睡衣', 0.27353), (`加厚外套', 0.26865), (`家居', 0.26277), (`床', 0.25459)] \cr\cr
    \textbf{Relevance Score:} \quad $0.343 \times 0.99965 + 0.13778 \times 0.99979 + 0.09616 \times 0.90725 + 0.10872 \times 0.99703 + 0.202 \times 0.88608 = 0.8553$ \cr
    \cr
    \hline
    \end{tabular}
\end{CJK*}
}
\caption{\small Two examples of the DeepBoW model. Both the query and the product are encoded to the synonym-expansion BoW representation. The relevance score can be calculated as shown in the table.}
\label{case}
\end{table*}

\section{Case Study}
\label{appx_case}
Table \ref{case} shows two examples of our DeepBoW model. Both query and product are encoded to the synonym-expansion BoW representation. The sparse BoW representation consists of a collection of word-weight pairs, which can be regarded as the bag-of-words with soft weight. The synonym-expansion representation can not only capture the importance of the words in the original text, but also incorporates pertinent synonymous terms. The relevance score can be calculated by aggregating the matching terms of the query's/product's sparse BoW representation.

These two examples show that our proposed sparse BoW representation has positive interpretability, signifying that the developer can analyze bad cases from the online search system and implement targeted optimizations. Furthermore, the developer can modify the terms in the sparse BoW representation directly to achieve the expected result. 
In a word, our DeepBoW model surpasses other deep relevance modeling approaches in terms of interpretability and flexibility, thereby rendering it eminently suitable for the e-commerce search system.

\end{document}